\documentclass[conference,10pt,a4paper]{IEEEtran}
\usepackage{graphicx, epstopdf}
\usepackage{amssymb, amsmath}
\usepackage{commath}
\usepackage{setspace}
\usepackage{acronym}
\usepackage{mathrsfs}
\usepackage{ifthen}
\usepackage{textcomp}
\usepackage{float}
\usepackage{subfigure}
\usepackage{color}
\usepackage{cite}
\IEEEoverridecommandlockouts

\begin{document}
\bstctlcite{ICC09_Ref2:BSTcontrol}

\title{Performance Analysis of Micro Unmanned Airborne Communication Relays for Cellular Networks}

\author{Weisi Guo, Conor Devine, Siyi Wang\textsuperscript{\dag}\\
School of Engineering, The University of Warwick, UK\\
\textsuperscript{\dag}Institute for Telecommunications Research, University of South Australia, Australia\\
Email: \{weisi.guo, conor.devine\}@warwick.ac.uk, \textsuperscript{\dag}siyi.wang@mymail.unisa.edu.au}

\IEEEoverridecommandlockouts
\maketitle

\begin{abstract}
This paper analyses the potential of utilising small unmanned-aerial-vehicles (SUAV) as wireless relays for assisting cellular network performance. Whilst high altitude wireless relays have been investigated over the past 2 decades, the new class of low cost SUAVs offers new possibilities for addressing local traffic imbalances and providing emergency coverage. We present field-test results from an SUAV test-bed in both urban and rural environments. The results show that trough-to-peak throughput improvements can be achieved for users in poor coverage zones. Furthermore, the paper reinforces the experimental study with large-scale network analysis using both stochastic geometry and multi-cell simulation results. 
\end{abstract}


\section{Introduction}

\subsection{Motivation}

Over the past decade, mobile traffic has been transformed from mainly voice-based to an amalgamation of different data types.  According to the latest report by the Cisco Visual Networking Index (VNI), the mobile video traffic will increase by \mbox{16-fold} over the next 5 years~\cite{Cisco12}, driven mainly by mobile video content.  Another key global trend is the human migration from rural to urban areas.  More than 50\% of the global population now live in cities, and this percentage is over 80\% for developed countries~\cite{UN13}.  Combining the mobile data growth and rapid urbanisation trends together, one can draw the conclusion that there will be an extremely high density of wireless communication links in cities.  Therefore, the battle ground for cellular communications success will lie in complex urban areas, where the signal propagation environment is very hostile.  In addition to the propagation challenge, the spatial- and temporal-traffic demand pattern is also extremely complex, with a large variance.  This has led to an inefficient utilisation of cellular network resources, with 50\% of the base stations (BSs) carrying over 90\% of the data at any given time.

Whilst the aforementioned challenges are especially prominent in cities, it is also a challenge in far reaching rural communities that seek greater information connectivity.  Traditional (Bell Labs~\cite{Claussen06}) and non-traditional wireless data operators (Google Loon Project and Facebook Drone Project~\cite{Facebook14}) are now examining the need to supplement fixed terrestrial wireless infrastructure with mobile airborne services.  Some of the technologies considered are high altitude long endurance (HALE) platforms, whilst others are autonomous small unmanned aerial vehicles (SUAVs).  This paper is primarily concerned with the latter.

\subsection{Heterogeneous Network Architecture}

\subsubsection{Benefits}

The mobile data traffic has approximately doubled every year for the past decade, and the wireless multiple access capacity also needs to increase at a similar rate.  To an extent, the challenges are being addressed by dynamic network elements that employ greater inter-cell cooperation, spectrum sharing, and load balancing techniques.  A promising solution that is being trumpeted by both industry and academia is small-cells, and they are widely seen as the solution to providing cost efficient capacity growth~\cite{Chu13}.  Small-cells can be femto-cells with a wired- or wireless-backhaul~\cite{Andrews08}, or wireless relays~\cite{Guo12JSACrelay}.  Almost all operators now include small-cells in their strategy to provide ubiquitous mobile broadband.  Typically, they are deployed in urban canyons (streets and alleyways) or in large indoor areas (shopping malls and business centres).
\begin{figure*}[t]
    \centering
    \includegraphics[width=0.95\linewidth]{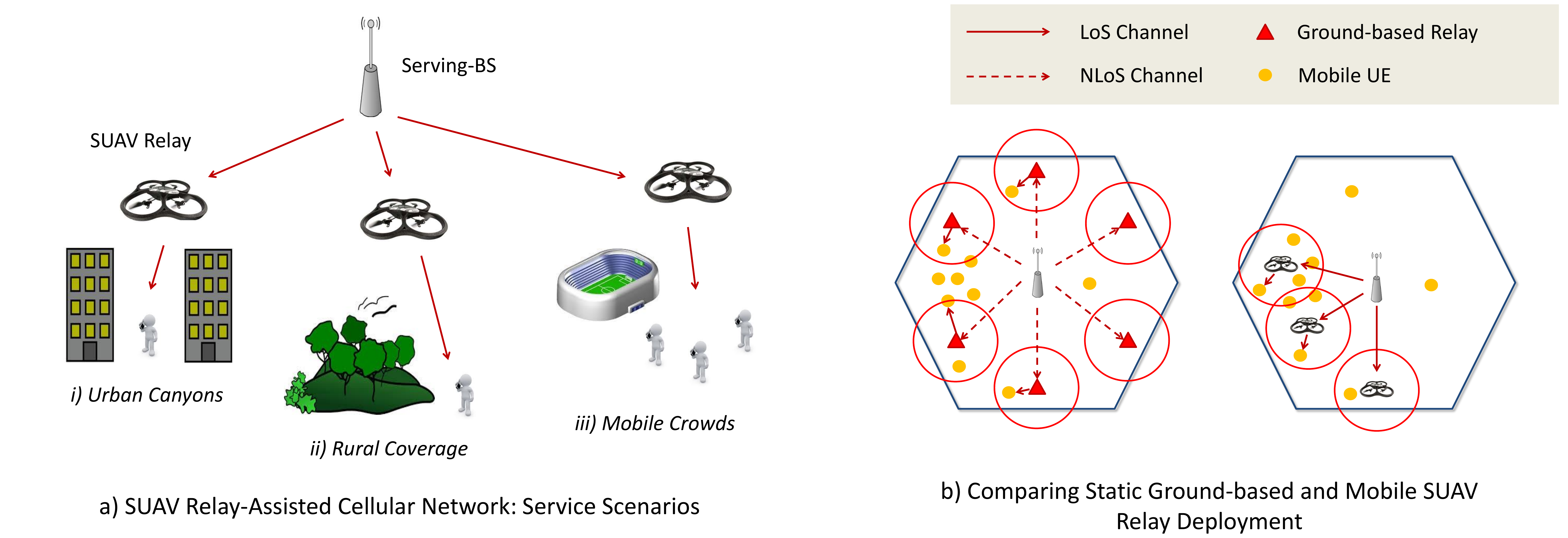}
    \caption{Illustration of: a) Service scenarios for SUAV relay-assisted cellular network; b) Static ground-based and Mobile SUAV relay deployment.}
    \label{System}
\end{figure*}

\subsubsection{Challenges}

Whilst the hardware of a small-cell (i.e., femto-cell) is relatively cheap, the site rental and backhaul costs are 10--20 folder greater.  Therefore, one of the challenges faced by operator deployed small-cells is their deployment location~\cite{Ge13, Guo13CM}.  The major drawback with \emph{static} deployments is that their service qualities (i.e., coverage area) has a very limited adaptation scope.  The rapidly changing nature of the traffic demand and the hostile urban propagation characteristics mean up to 20 small-cells are needed per BS, yielding a cell density explosion from 6 macro-cells per square kilometre to 100 heterogeneous cells per square kilometre for each operator.  The fine grain planning required for small-cells coupled with the increase in co-channel interference will yield a coverage planning complexity explosion.  These barriers have thus limited to proliferation of small-cells, as operators continue to carry out limited scale deployments, whilst keeping a watchful eye on the operational costs and increased interference issues~\cite{Guo12GreeNet}.

Airborne relaying offers some attractive benefits, in that the relays offer improved propagation channels to both the serving base station and the targeted users.  Furthermore, airborne relays are mobile and can dynamically target emerging traffic hotspots.  The operational costs of a large fleet of SUAVs are attractive as they do not require site or fixed backhaul rental costs.  Furthermore, their capital costs are in the same order of magnitude as small cells and can be highly autonomous in their operations.

\subsection{Paper Organisation}

The paper will first review existing literature in Section II, to give a background on aerial relay platforms and relay radio resource optimisation.  In Section III, a single SUAV relay test-bed will be introduced and it will be used to test the performance of a SUAV relay-assisted 3G network in urban canyon and rural areas.  In Section IV, a multiple SUAV relay-assisted 3G network is simulated in central London, and the mobile coverage benefits of SUAV relays are demonstrated.  A theoretical framework based on stochastic geometry is also presented to yield performance bounds.  The results in Section IV will also compare SUAV relay performance to several other load balancing and traffic management techniques such as dynamic sector re-configuration~\cite{Du03}.  Finally in Section V, the paper will examine aviation authority regulations for commercial UAV operations and the practicality of the concept.

\section{Terrestrial Mobile Relaying Review}

\subsection{High Altitude Relays}

Mobile relays and BSs have been studied and tested over the past 2 decades as a possible supplement and replacement to ground BSs.  High altitude ($\sim 20$~km) and large-scale ($\sim 1,000$~kg) UAV BSs and relays have been proposed in~\cite{Tozer01, Denton10}, and more recently by the Google Project Loon.  High altitude UAVs primarily seek to replace terrestrial BSs in areas of: poor access, low user density, poor ground-level propagation or very little existing infrastructure (i.e., wired backhaul, electricity grid).  In one or more of these scenarios, the financial justification for maintaining a high-altitude fleet of large-scale UAVs and ground piloting stations can outweigh a terrestrial network.  The main challenges here include accurate transmitter and receiver alignment, long-duration flight, and a cost-effective ground coverage radius.  In contrast, this paper reviews recent work on low-altitude SUAVs.  They provide a micro-scale mobile relay that attempts to provide two distinctive benefits: i) superior terrestrial propagation model, and ii) coverage for a small dense cluster of mobile user equipments (UEs).  Therefore, they are not to be confused with or compared to high-altitude UAVs directly.

\subsection{Low Altitude Relays}

Indeed, several mobile relaying concepts have been proposed in the past few years.  Mobile relaying does not have to create new network entities such as new relay nodes, but can in fact opportunistically leverage on existing objects in the world such as: i) public transport vehicles~\cite{Panos12}, and ii) other mobile UEs~\cite{Scaglione06, Chatzigeorgiou08ISCCSP}.  The relaying protocol can either be of a cooperative or non-cooperative nature.  Due to the opportunistic nature of the aforementioned mobile relaying techniques, the performance tend to suit \emph{delay-tolerant} data, or users that are in \emph{poor coverage} areas.

In order to improve the delivery performance to all data types, dedicated hardware is required for mobile relaying (non-opportunistic).  On a low altitude level, robotic BSs that can relocate itself on roof-tops has been conceptually proposed in~\cite{Claussen06} to address the shifting urban traffic patterns.  To achieve this practically, the emergence of affordable low-power and -weight SUAVs ($\sim 10$~kg) is an attractive host for mobile relays.  Detailed analysis of it acting in a multi-cell cellular network environment has been lacking until recently.  In the past 2 years, several papers have exploited the idea that flying relays can dynamically follow large crowds of users and provide improved coverage in either emergency situations~\cite{Zhao12} or during traffic overload in the air interface of the network~\cite{Rohde12}.

Fig.~\ref{System}a illustrates a number of environments that can potentially benefit from relay assistance~\cite{Guo12JSACrelay, Rohde12}.  The scenarios that this paper will consider are: i) urban canyons (i.e., alleyways, indoors), rural areas, and mobile crowds.  Fig.~\ref{System}b illustrates the concept that static ground-based relays must be deployed to serve areas that have a long-term need for performance improvement, such as near the edge of the serving-BS.  They cannot therefore address the emergence of an UE hotspot effectively.  On the other-hand, mobile relays can converge on an emerging hotspot and provide effective coverage to improve link level spectral efficiency.

In terms of radio-resource assignment, multi-cell relaying research in~\cite{Guo12JSACrelay} has shown that co-frequency radio resource allocation between the relay and direct communication channels will yield the greatest overall network throughput.  In terms of optimal relay deployment location, multi-cell relaying research in~\cite{Guo12JSACrelay, Guo13CM, Rohde12} has shown that considering the multi-cell interference environment is important.  Solutions which only consider the serving signal power will deliver sub-optimal and sometimes degrading performances to the overall network.  By considering the effects of interference when relaying data, significant throughput improvements can be made.
\begin{figure}[t]
    \centering
    \includegraphics[width=0.95\linewidth]{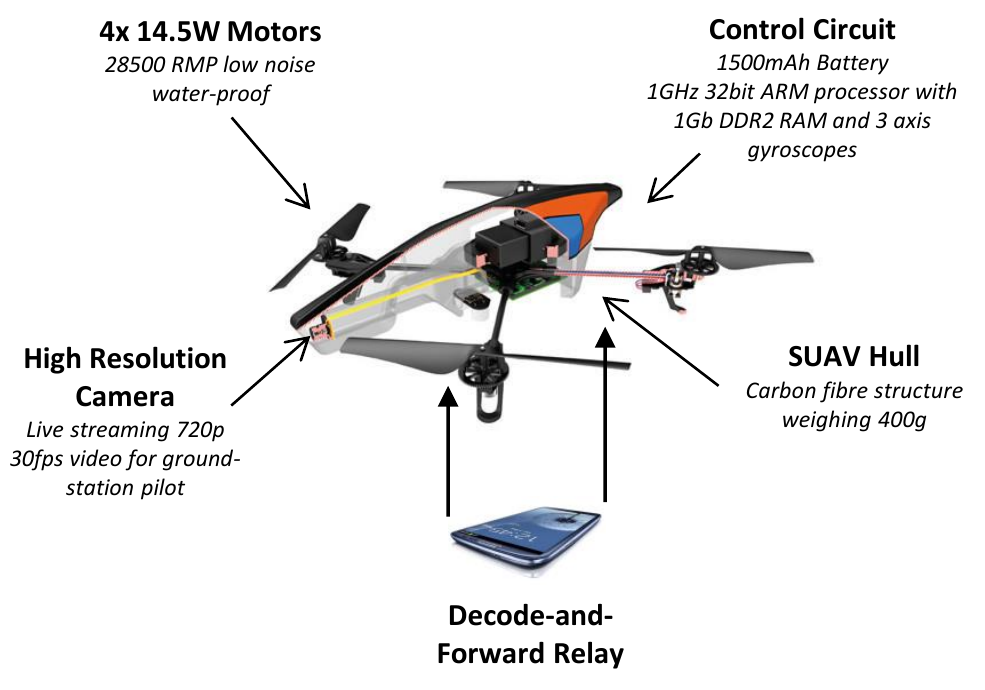}
    \caption{SUAV wireless relay test-bed that consists of a Parrot AR Drone 2.0 with an UE acting as a wireless relay.}
    \label{Testbed}
\end{figure}

\section{Single UAV Relay Field-Test Results}

\subsection{Hardware}

The paper first shows results of a single UAV relay node (RN) serving a single UE in poor coverage areas (urban outdoor, urban indoor, and rural outdoors).  The configuration of the test-bed, the experimental setup, and the performance gains will be presented.  The test-bed employs a Parrot AR Drone Mk2, which is a lighter model of the test-bed developed in~\cite{Rohde12}.  The drone is piloted by another user who does not take part in the throughput experimentation. The drone itself weighs approximately 400~g and has dimensions 0.51~m $\times$ 0.45~m.  It is equipped with four 14.5~W rotors and has a radio communications range of 50 metres.  As shown in Fig.~\ref{Testbed}, the drone is fitted with a wireless RN on its undercarriage.  The RN communicates with the serving BS on the 3G channel, and employs the Decode-and-Forward (DF) protocol to relay the signal to and from a test UE.  The specific hardware used in this test-bed is a UE in client mode.  The BS-RN channel is a 3G in-band channel, and the RN-UE channel is a out-of-band channel using the Wi-Fi transmission band.  The throughput test is performed on a test mobile-phone which is permanently connected to the drone's RN.  The drone's RN is connected to the nearest serving-BS and relays data continuously to the test UE.
\begin{figure}[t]
    \centering
    \includegraphics[width=0.95\linewidth]{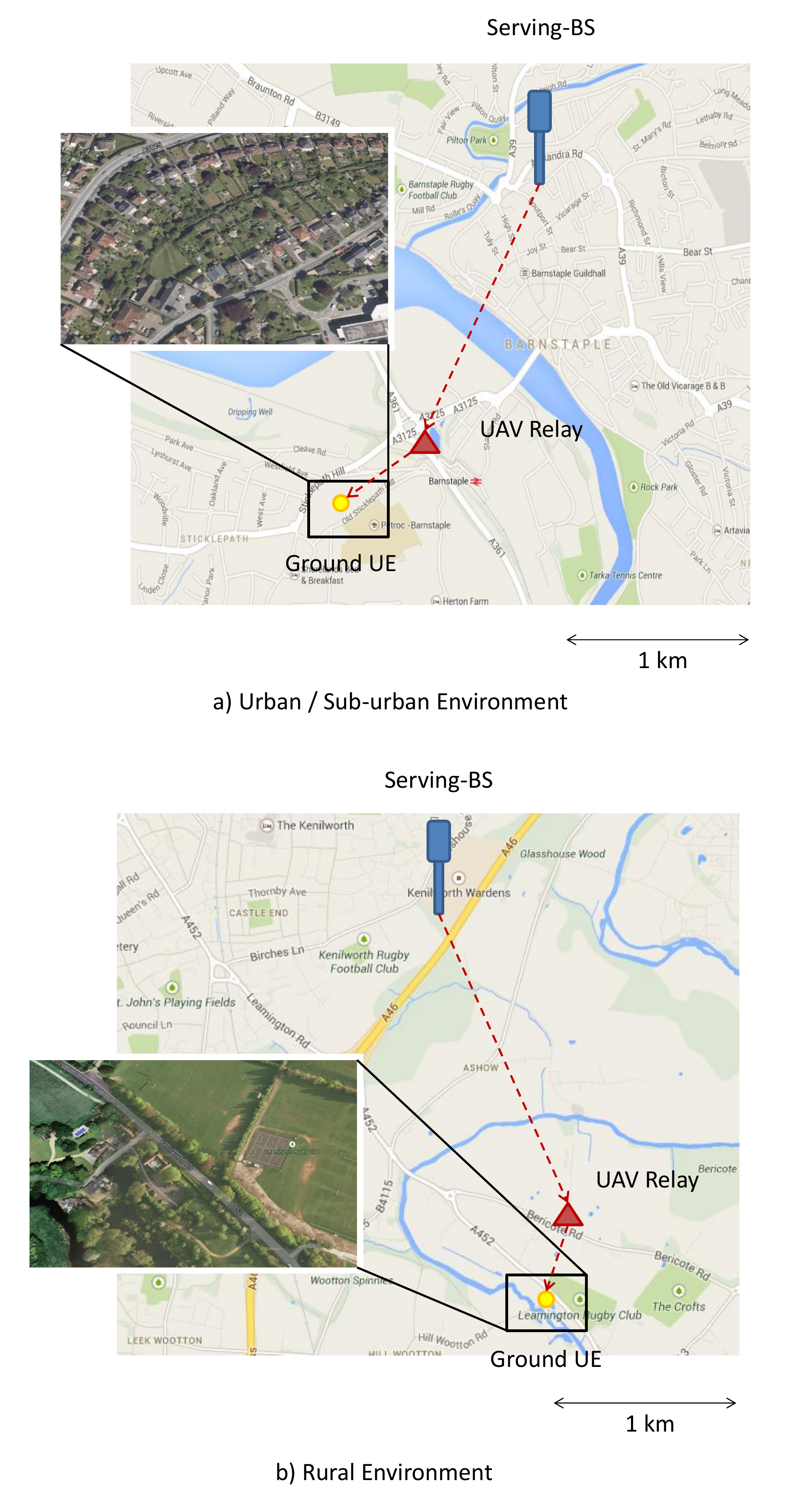}
    \caption{Map of urban and rural environments for experimental UAV RN testing.}
    \label{Map}
\end{figure}

\subsection{Urban and Rural Results}

\begin{figure}[t]
    \centering
    \includegraphics[width=0.95\linewidth]{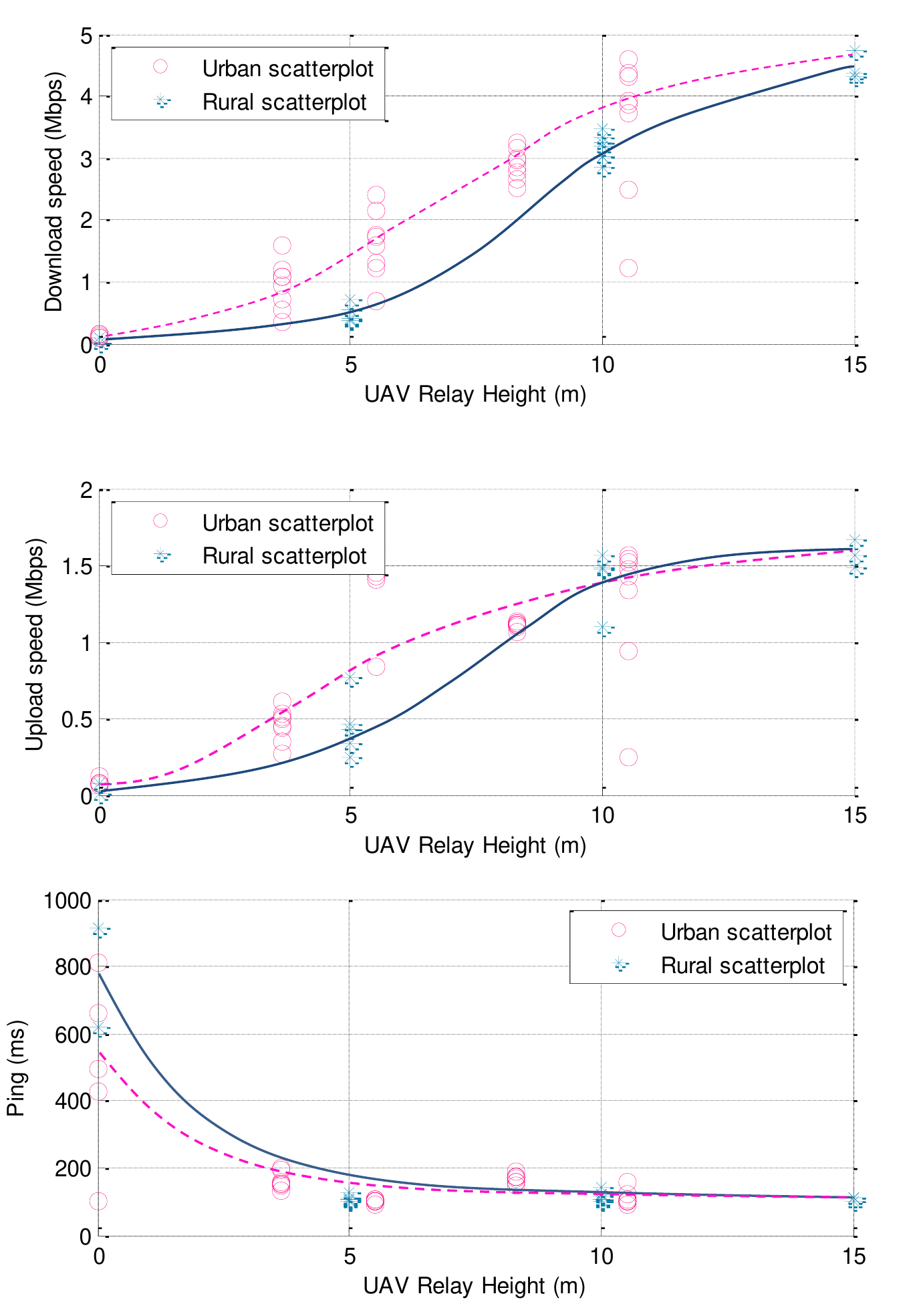}
    \caption{SUAV 3G RN test-bed results at different RN altitudes and at different locations for: downlink throughput, uplink throughput, and ping time.  Symbols show accumulated test data results over different locations and times, and lines show the average of the results.}
    \label{Results1}
\end{figure}

Test-bed results were performed in several urban and rural locations which do not have a strong 3G data coverage outdoors.  Fig.~\ref{Map} shows two example locations of an urban and rural setting, with the location of the serving-BS and the test point.  The combined urban and rural results are presented in Fig.~\ref{Results1}, which shows the downlink throughput, uplink throughput, and ping time for different relaying altitudes.  The general observation is that the two sets of throughput results show a very similar monotonically increasing trend with RN height, which saturates at approximately an altitude of 15 metres.  The surround building and foliage did not exceed 15 metres, so it is conceivable that higher altitudes are required for alternative terrains.  The conditions during experimental testing were also very windy, and this affected the control of the SUAV more than it did the results, because the performance of the RN is more affected by its altitude than horizontal position.

The results from numerous test runs at different locations and at different times, it was found that the downlink throughput improvement can be increased from 150~kbps to a saturation of 5~Mbps, and the uplink throughput improvement can be increased from 70~kbps to a saturation of 1.5~Mbps.  The ping time results show a monotonically decreasing trend with RN height, which saturates at approximately an altitude of 5 metres at 100~ms.  A similar test was performed for indoor UEs in an urban area with no data coverage.  The UAV RN was made to hover near a window.  The performance result achieved are a downlink throughput improvement to 2 Mbps and an uplink throughput improvement to 1 Mbps.

In summary, the wireless RN has achieved a trough-to-peak throughput and ping time improvement for UEs in poor coverage urban and rural areas.  Thus far, the paper reviewed the results of a single UAV RN serving a single UE in a variety of poor coverage areas.  The paper now presents both theoretical and large-scale simulation results related to large scale UAV relaying deployment in a multi-cell cellular network.

\section{Multiple UAV Relay Results}

\subsection{Theoretical Capacity Gain using Stochastic Geometry}

As illustrated in Fig.~\ref{System}b, the fixed ground RNs generally have Line-of-Sight (LoS) to the serving-BS and Non-Line-of-Sight (NLoS) to the targeted UEs~\cite{80216m}.  The UEs generally do not have LoS to the serving-BS.  The interference from other co-channel cells (RN or otherwise) generally have a Non-Line-of-Sight (NLoS) channel.  The \emph{propagation pathloss model} is statistically modelled on the 3GPP urban micro scenario~\cite{3GPP10}:
\begin{equation}\begin{split}
\label{Pathloss}
\Lambda = K d^{-\alpha} = \left\{
\begin{array}{l l}
\alpha = 2   &  \mbox{for LoS} \\
\alpha = 4   &  \mbox{for NLoS} \\
\end{array} \right.,
\end{split}\end{equation}
and the constant $K$ is dependent on the transmission frequency and is approximately $10^{-4}$ for 2.1 GHz.  Therefore, the advantage of the mobile SUAV RN is therefore two fold: i) gain LoS with the targeted UEs and the serving BS, ii) can spatially-track the UE hotspots.

A theoretical performance gain based on the previously mentioned propagation channels in \eqref{Pathloss} can be found using stochastic geometry~\cite{Haenggi12, Andrews09}.  By only examining UEs that are connected to RNs and assuming that the BS-RN channel is not the limiting channel, useful performance bounds can be found.  Referring to the multi-cell model of a cellular network, there exists a closed form expression for the complementary cumulative-distribution-function (CCDF) of the Signal-to-Interference Ratio (SIR) at a fixed point $r$ away from the RN:
\begin{equation}\begin{split}
\label{CCDF}
P(\gamma_{r} > \xi) = \left\{
\begin{array}{l l}
\exp \left[ -\lambda \pi r\sqrt{\xi}\arctan\left(\frac{\sqrt{\xi}}{r}\right) \right]   &  \mbox{SUAV RN} \\
\exp \left[ -\lambda \pi r^{2} \sqrt{\xi}\arctan\left(\sqrt{\xi}\right) \right]   &  \mbox{Ground RN} \\
\end{array} \right.
\end{split}\end{equation} where $\lambda$ is the RN density.  By averaging across all distances $r$, the mean network SIR, as well as metrics such as outage and capacity can be found for each case.
\begin{figure}[t]
    \centering
    \includegraphics[width=0.95\linewidth]{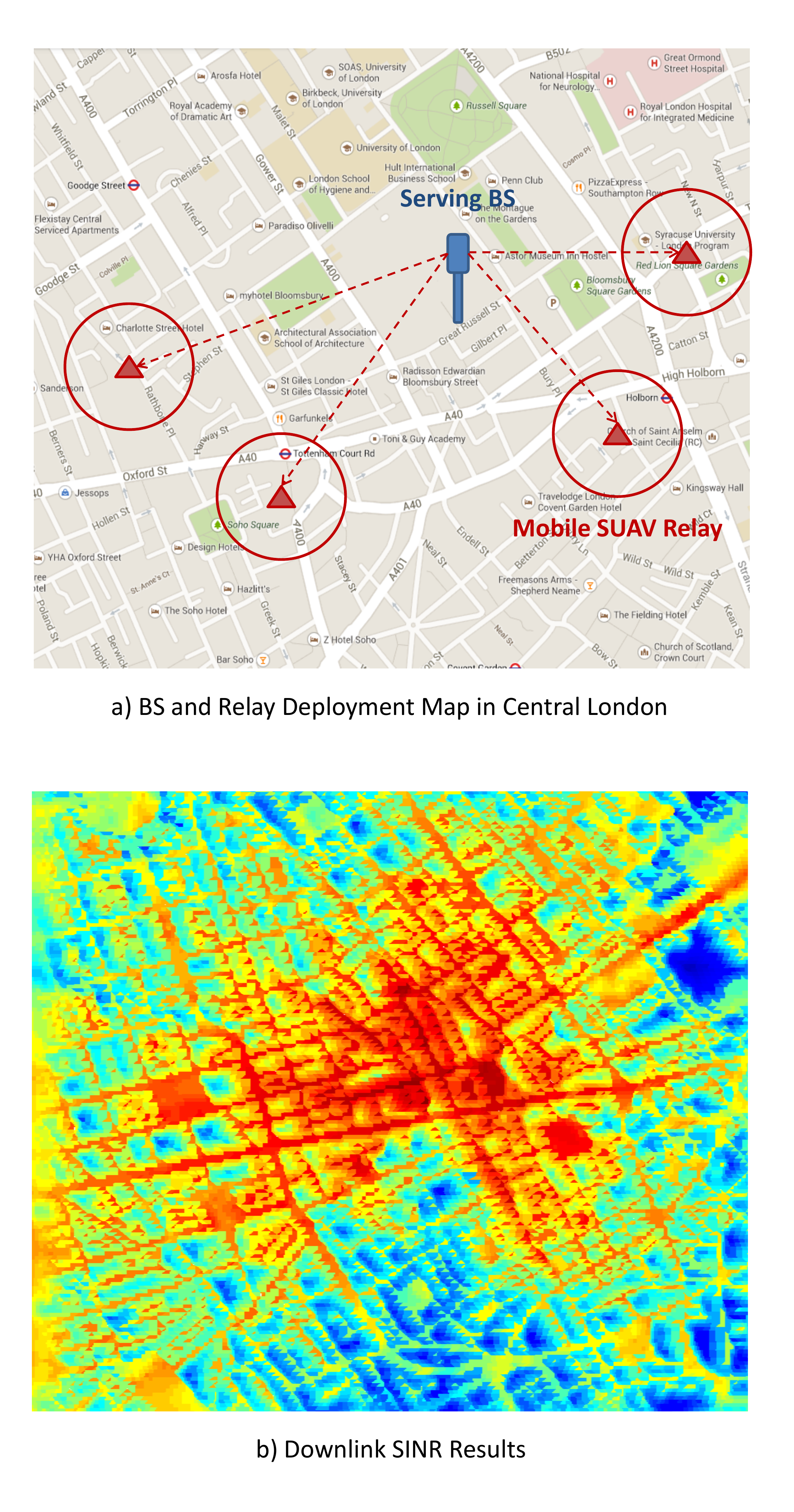}
    \caption{Central London SUAV RN deployment scenario with an illustration of an example BS: a) map of coverage area (2~km $\times$ 2~km) with BS location and snap-shot of SUAV location, b) resulting downlink SINR performance.}
    \label{Map2}
\end{figure}
\begin{figure}[t]
    \centering
    \includegraphics[width=0.95\linewidth]{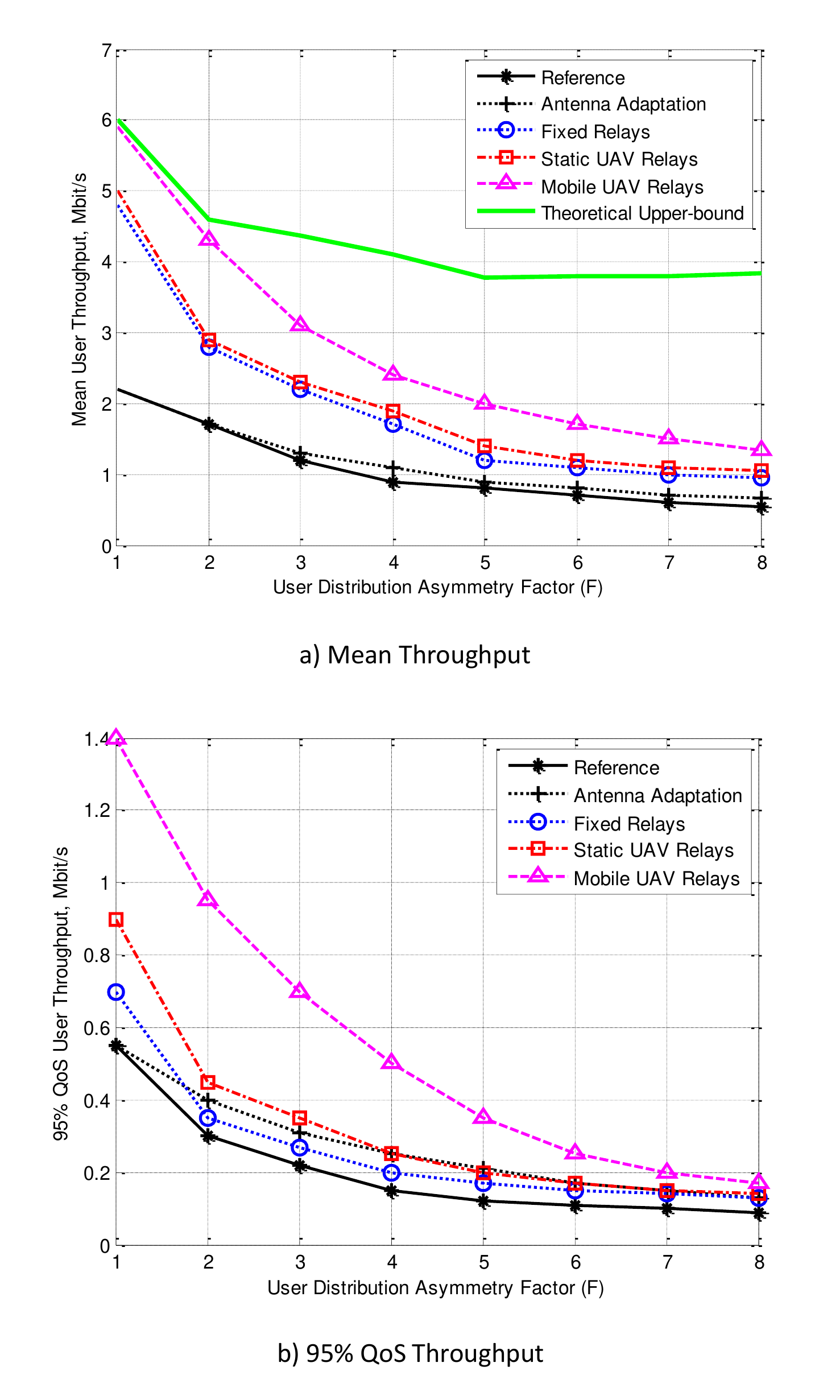}
    \caption{Mean throughput and QoS results for different schemes.}
    \label{Results2}
\end{figure}

\subsection{London City Simulation Results}

The simulation results analyse the performance of several cellular techniques against the asymmetry of traffic distribution.  The setting is based in 5~km $\times$ 9~km central London with real network BS locations and ray-traced pathloss modelling using industrial packages: PACE 3G and Forsk Atoll, as well as an industrially bench-marked proprietary simulation software VCEsim~\cite{Guo12JSACrelay}.  An example of the experiment for a single BS is illustrated in Fig.~\ref{Map2}, which shows a BS with a number of SUAV RNs serving traffic hotspots.  The concept of traffic distribution asymmetry factor ($F$) is defined as the ratio between the volume of traffic (data demanding users) in the serving-BS and the mean traffic in all BSs. As the asymmetry factor $F$ increases, the mean throughput experienced by each UE in the serving-BS decreases due to the reduced number of radio resource blocks per UE.

For a reference configuration with no RNs (homogeneous network), the results in Fig.~\ref{Results2} show that a peak mean user throughput of 2.2~Mbps and 95\% of users (QoS) can achieve at least 0.5~Mbps, decaying with a half-life of $F=3$.  The theoretical upper-bound considers a network with no interference and the capacity of each user scales only with the radio resources available.  The load balancing technique~\cite{Du03} can achieve an 11\% improved mean throughput and 46\% improved QoS over the reference.  For a heterogeneous network with 6 fixed ground wireless RNs per BS, the mean throughput improvement is 65--120\% and the QoS improvement is 20--40\% over the reference.  The greatest improvement is achieved for a uniform distribution of users, which means fixed ground RNs are no better placed in dealing with unpredictable surges of traffic demand.

For a heterogeneous network with 6 mobile wireless UAV RNs per BS, the mean throughput improvement is 160--190\% and the QoS improvement is 140--210\% over the reference.  Compared to fixed relaying, the mean throughput improvement is 22--60\% and the QoS improvement is 70--180\%.  The greatest improvement is achieved for high user asymmetry patterns ($F \geqslant 3$), showing that mobile UAVs are more suitable for non-uniform traffic patterns.

\section{Discussion on Regulations \& Practicality}

Whilst the operational laws for UAVs differ in each country and many laws are being re-written at the time of this article, we give some general insight into current laws.  The Civil Aviation Authority (CAA) in the United Kingdom (UK) gives legal guidance on the usage of \emph{Unmanned Aircraft System Operations in UK Airspace}~\cite{CAA10}.  The regulation is dependent on the weight class of the UAV, of which for the purpose of light-weight ($<20$~kg) UAVs, there are no requirements for airworthiness and registration.  Permission and pilot qualification is needed if the purpose of the aircraft is to conduct commercial operations or flown in congested areas.  In terms of operational range for small-sized UAVs, visual-LoS (VLoS) is acceptable without permission.  This defined as a maximum distance of 0.5~km horizontally and 100~m in altitude from the remote pilot.  For extended- and beyond-VLoS, the UAV must be equipped with means to avoid collisions, be visible to other airspace users, resilient to relevant meteorological conditions, and suitable range control.

In realistic commercial operations of the UAV RN, the current test-bed cannot meet the air authority conditions.  A larger UAV that can achieve collision avoidance and be resilient to the effects of wind is needed.  Currently this is achievable within the $<20$~kg category and therefore, given the correct permissions and automated piloting softwares, it is feasible that UAV RNs can be automatically associated with cellular network BSs in the near future.

\section{Conclusions}

Over the past decade, mobile data growth has changed the way wireless communication networks think about deployment and operations.  Whilst this is especially prominent in cities, where most of the world's population now live, it is also a challenge in far reaching rural communities that seek greater information connectivity.  Traditional and non-traditional wireless data operators (i.e., Google and Facebook) are now examining the need to expand their fixed wireless infrastructure with mobile airborne services.

Low-altitude SUAVs provide a micro-scale mobile communications relay that can attempt to provide two distinctive benefits: i) superior propagation model, and ii) provide increased bandwidth reuse for emerging traffic hotspots.  Therefore, they are not to be confused with or directly compared with high-altitude UAV technologies.  The paper presents experimental field-test results based on a 3G SUAV relay test-bed, as well as simulation and theoretical bounds to performances.  The results in both rural and urban environments show a similar trough-to-peak throughput and ping time improvements.  The paper goes on to compares the SUAV relay's performance with alternative load balancing and static relaying techniques.  Overall, the improvement over existing methods in both mean throughput ($>$22\%) and QoS ($>$70\%) is significant enough to seriously consider the technology. \\

\section*{Acknowledgment}
The work in this paper is partly supported by the University of Warwick's URSS Scheme.

\bibliographystyle{IEEEtran}
\bibliography{IEEEabrv,Mag}

\end{document}